 \documentclass[twocolumn,aps,prc,showpacs,superscriptaddress,preprintnumbers,floatfix,nofootinbib]{revtex4}
\usepackage{epsfig,graphics}
\usepackage{graphicx}
\usepackage{dcolumn}
\usepackage{bm}
\usepackage{amsmath}
\usepackage[usenames]{color}
\usepackage{ulem} 
\usepackage[colorlinks,linkcolor=blue,urlcolor=blue,citecolor=blue]{hyperref}
\usepackage{hyperref}
\usepackage{CJK}

\begin{document}
\begin{CJK*}{UTF8}{gbsn}

 \title{Testing the collectivity in large and small colliding systems with test particles}
\author{Han-Sheng Wang}
\affiliation{Key Laboratory of Nuclear Physics and Ion-beam Application (MOE), Institute of Modern Physics, Fudan University, Shanghai 200433, China}
\affiliation{Shanghai Research Center for Theoretical Nuclear Physics, NSFC and Fudan University, Shanghai $200438$, China}
\author{Guo-Liang Ma}
\email[]{glma@fudan.edu.cn}
\affiliation{Key Laboratory of Nuclear Physics and Ion-beam Application (MOE), Institute of Modern Physics, Fudan University, Shanghai 200433, China}
\affiliation{Shanghai Research Center for Theoretical Nuclear Physics, NSFC and Fudan University, Shanghai $200438$, China}


\begin{abstract}
We propose a test-particle method to probe the transport dynamics of the establishment and development of collective flow in large and small systems of heavy-ion collisions. We place test particles as passengers into the partonic medium created by Au$+$Au midcentral collisions at $\sqrt{s_{NN}}$ = 200 GeV and $p$ $+$ Pb central collisions at $\sqrt{s_{NN}}$ = 5.02 TeV, using a multiphase transport model. With the help of test particles in two extreme test cases, we demonstrate that parton collisions play an important role in establishing and developing collectivity in large and small colliding systems. The collectivity established by final state parton collisions is much stronger in large colliding systems compared to small colliding systems. The collectivity from the initial state can persist or survive more easily in small colliding systems than in large colliding systems due to fewer parton collisions. Our study provides a new method to understand the origin of collectivity in large and small colliding systems at the BNL Relativistic Heavy Ion Collider and the CERN Large Hadron Collider.
\end{abstract}

\pacs{}

\maketitle

\section{INTRODUCTION}
\label{introduction}

The ultrarelativistic heavy-ion collisions at the BNL Relativistic Heavy Ion Collider (RHIC) and the CERN Large Hadron Collider (LHC) are believed to create a fireball of quark-gluon plasma (QGP) at extreme conditions of pressure and energy densities, through colliding two nuclei with apposite velocities closing to the speed of light \cite{STAR:2005gfr,PHENIX:2004vcz,ALICE:2008ngc,Bzdak:2019pkr,Luo:2017faz,Xu:2014tda}. High pressure gradient drives the hydrodynamic expansion of the produced fireball, which leads to strong collective flow, including both radial and anisotropic flow~\cite{Siemens:1978pb,Kolb:2000sd,Teaney:2000cw,Song:2007ux,Jeon:2015dfa,Shen:2020mgh,Lao:2017dtr,Waqas:2020ioh}. Due to the elliptic geometry for the noncentral collision or the initial energy density fluctuations in the overlapped zone, the spatial asymmetry of the strongly interacting quark-gluon plasma (sQGP) can be generated~\cite{Gale:2013da,Alver:2010gr,Ma:2010dv,Ma:2016hkg}. This spatial asymmetry can be translated into momentum anisotropy of the final particles through pressure gradients in hydrodynamics, i.e., the formation of collective flow~\cite{Ollitrault:1992bk,Heinz:2013th,Qin:2010pf}. Therefore, collective flow is considered an important probe of the hydrodynamic behavior of the QGP~\cite{Stoecker:2004qu,Yan:2017ivm,Lan:2022rrc}.
	
The collective flow in large systems created by heavy-ion collisions ($A + A$) has been well explained by hydrodynamic models. But the measurements of collective flow in small systems, such as $p + p$ and $p$ $+$ Pb collisions at the LHC \cite{CMS:2010ifv,CMS:2012qk,ALICE:2012eyl,ATLAS:2012cix} and $d$ $+$ Au collisions at RHIC \cite{PHENIX:2014fnc,STAR:2015kak}, aroused some debate about the paradigm of collective flow~\cite{Dusling:2015gta,Loizides:2016tew,Nagle:2018nvi}. Many theoretical efforts have been made to understand the origin of collective flow in small colliding systems. It can be basically classified into two categories depending on whether the origin comes from the initial or final state. The initial state of color glass condensate (CGC) has also been proposed as a possible mechanism, contributing to the experimentally measured `flow' in small colliding systems~\cite{Dumitru:2010iy,Dusling:2013oia,Skokov:2014tka,Schenke:2015aqa,Schlichting:2016sqo,Kovner:2016jfp,Iancu:2017fzn,Mace:2018vwq,Nie:2019swk,Shi:2020djm}. The final state of hydrodynamics also can transform the initial geometric asymmetry into the final momentum anisotropic flow through the pressure gradient of the QGP~\cite{ Bozek:2011if,Bzdak:2013zma,Shuryak:2013ke,Qin:2013bha,Bozek:2013uha,Bozek:2015swa,Song:2017wtw}. However, the applicability of hydrodynamics to small colliding systems is still questionable since the Knudsen number (the ratio of micro to macro distance/time scales) is not small in small colliding systems which could result in a strong deviation from local thermal equilibrium~\cite{Yan:2017ivm,Niemi:2014wta,Heinz:2019dbd}. In principle, when the Knudsen number $\gg 1$, statistical mechanics should be applied.  A parton transport model, a multi-phase transport (AMPT) model \cite{Lin:2004en} also described the experimental data in both large and small systems \cite{Bzdak:2014dia,OrjuelaKoop:2015jss,Ma:2016bbw}. Due to the fact that the majority of partons have no scatterings for small colliding systems at RHIC and the LHC, a parton escape mechanism has been proposed to explain the formation of azimuthal anisotropic flow \cite{He:2015hfa,Lin:2015ucn}, which found that anisotropic parton escape dominates the flow generation in small colliding systems. Meanwhile, parton collisions have been shown to be crucial for generating anisotropic flow~\cite{Ma:2016bbw,Ma:2021ror}. It is important to understand the nature of flow in parton escape mechanism, because the transport model is considered beneficial for systems with insufficient multiplicity, and it bridges the gap between nonequilibrium and equilibrium states~\cite{Schlichting:2016sqo,Berges:2020fwq, Romatschke:2017ejr,Schlichting:2019abc}. Therefore, studying flow by parton transport models is expected to provide important information about the nature of collectivity.

In this work, we propose a test-particle method to investigate the relationship between collective flow and parton collisions. The parton collisional effects on different settings of systems will be compared to show the nature of collectivity in both large and small colliding systems. Our paper is organized as follows. In Sec. II, we introduce our model,  propose a test-particle method, and define our observables. In Sec. III, we present our main results to discuss the parton collisional effects on the collectivity in large and small colliding systems. Finally, we summarize in Sec. IV.

\section{MODEL AND METHOD}
\label{results}
 \subsection{A multiphase transport model}
\label{AMPT}

The string melting version of AMPT model consists of four main stages of heavy-ion collisions, i.e., initial state, parton cascade, hadronization, and hadronic rescatterings. The initial state with fluctuating initial conditions is generated by the heavy ion jet interaction generator (HIJING) model~\cite{Gyulassy:1994ew}. In HIJING model, minijet partons and excited strings are produced by hard processes and soft processes, respectively. In the string melting mechanism, all excited hadronic strings in the overlap volume are converted to partons according to the flavor and spin structures of their valence quarks \cite{Lin:2001zk}. The partons are generated by string melting after a formation time,
\begin{equation}
	t_f=E_{H}/m^2_{T,H},
	\label{tf }
\end{equation}
where $E_{H}$ and $m_{T,H}$ represent the energy and transverse mass of the parent hadron. The initial positions of partons from melted strings are calculated from those of their parent hadrons using straight-line trajectories. 
The interactions among partons are described by the Zhang's parton cascade (ZPC) parton cascade model \cite{Zhang:1997ej}, which includes only two-body parton elastic scattering  with a $gg \rightarrow gg$ cross section:
\begin{equation}\label{sigma}
\frac{d\sigma }{d\hat{t}}=\frac{9\pi \alpha ^{2}_{s}}{2}(1+\frac{\mu^2 }{\hat{s}})\frac{1}{(\hat{t}-\mu^2)^2},
\end{equation}
where $\alpha_s$ is the strong coupling constant (taken as 0.33), while $\hat{s}$ and $\hat{t}$ are the usual Mandelstam variables. The effective screening mass $\mu$ is taken as a parameter in ZPC for adjusting the parton interaction cross section. It is set as 2.265 fm$^{-1}$ leading to a total cross section ($\sigma$) of about 3 mb for elastic scattering. The previous studies have shown that a parton interaction cross section of 3 mb can well describe both large and small colliding systems at RHIC and the LHC~\cite{Lin:2014tya,OrjuelaKoop:2015jss,Ma:2016bbw,Ma:2016fve,He:2017tla,Lin:2021mdn}. Meanwhile, a quark coalescence model is used for hadronization at the freeze-out of parton system. The final-state hadronic scatterings in the hadronic phase are simulated by a relativistic transport (ART) model \cite{Li:1995pra}.  Since we are only interested in the parton collisional effect on the evolution of collective flow, we will focus on the stage of parton cascade and ignore the effects from hadronization and hadronic rescatterings in this study. 

In our convention, the $x$ axis is chosen along the direction of the impact parameter $b$ from the target center to the projectile center, the $z$ axis is along the projectile direction, and the $y$ axis is perpendicular to both the $x$ and $z$ directions. The time $t$ starts when the two nuclei are fully overlapped in the longitudinal direction.

\subsection{Test particle method}
\label{METHODS}

\begin{figure}[htbp]
	\centering
	\includegraphics[width=0.9\linewidth]{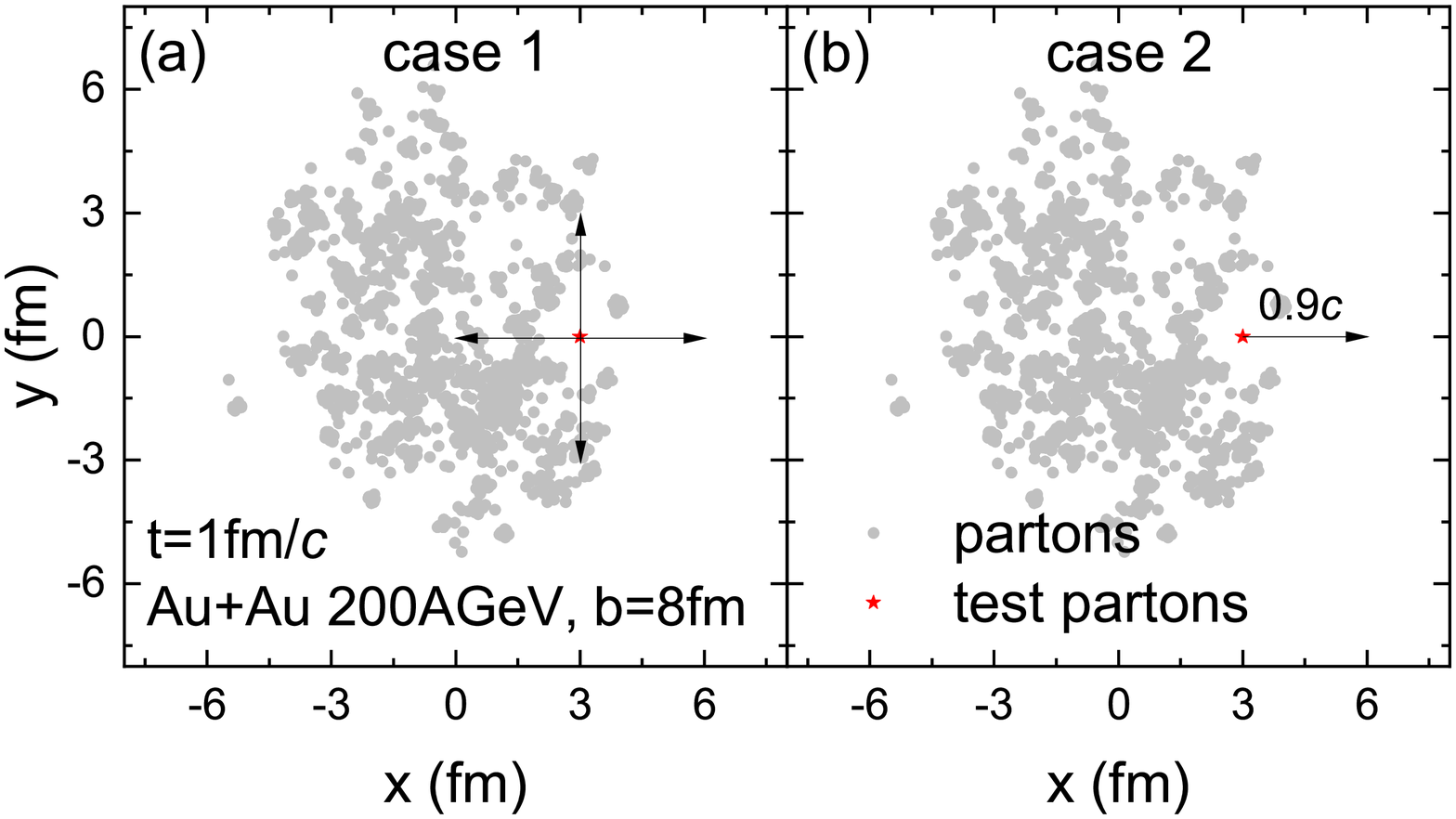}
	\caption{(Color	online) The two cases of transverse view of parton system (within $|\eta_{s}| < $0.5) at $t = 1$ fm/\textit{c} in a Au $+$ Au midcentral collision event at 200 GeV from the AMPT model. The positions of medium partons are represented by solid gray circles. The positions of four test particles are represented by red stars. The initial velocities of four test particles are shown by thin black arrows.}
	\label{test_particle_statistics}
\end{figure}

In order to study the relationship between parton collisions and collective flow in the expanding fireball, several test particles (partons) are added to the original parton system at the stage of parton cascade. Like placing leaves in a stream, the test particles are expected to follow and reflect the collective motion of the entire system due to their interactions with the medium. The test particles and medium behave like passengers and carriers, respectively. We set up two cases of scenes:  
Case 1: The four test particles are placed with the different initial velocities $\vec{v}_1=(0.9 c, 0, 0)$, $\vec{v}_2=(-0.9 c, 0, 0)$, $\vec{v}_3=(0, 0.9 c, 0)$, and $\vec{v}_4=(0, -0.9 c, 0)$, respectively, where \textit{c} is the speed of light, as shown in Fig.~\ref{test_particle_statistics}(a).
Case 2, The four test particles are placed with a same initial velocity $\vec{v}=(0.9 c, 0, 0)$, as shown in Fig.~\ref{test_particle_statistics}(b). Case 1 corresponds to the case where the test particles have no initial collective flow due to their isotropic initial velocities. However, case 2 corresponds to the case where the test particles have an initial collective flow due to their common initial velocities. Note that the initial collective flow is only introduced to the test particles for case 2, which is different from our previous study where the initial collectivity was introduced to all particles of the system~\cite{Nie:2019swk}. If we consider that the mass of the test particle is $0.05$ GeV (since we always take the test particles as light quarks), with the initial velocity $\beta$ the transverse momentum $p_T=\beta/(\sqrt{1-\beta^2}) m_q \approx $ 0.1 GeV/\textit{c}. 
We will put four test particles at the position of $ \vec{r}_{0}$ at the time of $t_0$ in midcentral Au$+$Au collisions at $\sqrt{s_{NN}}$ = 200 GeV and central $p$ $+$ Pb collisions at $\sqrt{s_{NN}}$ = 5.02 TeV. For Au$+$Au collisions, we set $\vec{r}_{0}$ = (3, 0, 0 fm) and $t_0$ = 1 fm/\textit{c}, while for $p$ $+$ Pb collisions, $\vec{r}_{0}$ = (2, 0, 0 fm) and $t_0$ = 1 fm/\textit{c}, unless specified otherwise. Using the AMPT model, we simulated $10^4$ events of Au$+$Au midcentral collisions ($b = 8$ fm) at 200 GeV for cases 1 and 2, and $10^4$ events of $p$ $+$ Pb central collisions ($b = 0$ fm) at 5.02 TeV for cases 1 and 2. With the help of the test particles, we focus on the time evolution of the collectivity of partonic matter for cases 1 and 2 in large and small colliding systems. The four test cases can be witnessed by the animations at~\cite{cartoon}. Note that in our simulation, the medium particle will change its momentum after colliding with a test particle. However, we have checked that the results are similar regardless of whether the medium particle changes its momentum or not.

\subsection{Test particle observables}
\label{obser}

\begin{table*}[htbp]
	\caption{The expectations of the eight defined observables in the initial states of cases 1 and 2. }. 
	\label{table:obsopp}
	\centering
	\begin{tabular}{p{150pt}p{40pt}p{40pt}p{50pt}p{50pt}p{25pt}p{25pt}p{30pt}p{30pt}}
	\hline
	\hline
	 &$\left \langle D|R|\right \rangle$ &$\left \langle D|v|\right \rangle$& $\left \langle D\cos{\Delta\phi}\right \rangle$&$\left \langle D\cos{\Delta\phi_v}\right \rangle$& $\left \langle N_r \right \rangle $   &$\left \langle N_v \right \rangle$& $\left \langle \cos{\phi}\right \rangle$&$\left \langle \cos{\phi_v}\right \rangle$\\
	\hline
	case 1 (zero collectivity) &0&1.45&-1.0&1/3&1.0&0  &1.0&0\\
	\\
	case 2 (strongest collectivity)      &0&0     &-1.0&-1.0 &1.0&1.0&1.0&1.0\\
	\hline
	\hline
	\end{tabular}
	\end{table*}

To investigate the collective motion of test particles, we first need to define several observables to probe collectivity. The dispersities of relative position and velocity of $N$ test particles are defined as follows:
\begin{eqnarray}\label{magnitude}
\left \langle D|R|\right \rangle  & = & \frac{1}{N(N-1)}\sum ^N _j \sum ^N _{i\neq j}|\vec{r}_i-\vec{r}_j|,  \notag \\
\left \langle D|v|\right \rangle  & = & \frac{1}{N(N-1)}\sum ^N _j \sum ^N _{i\neq j}|\vec{v}_i-\vec{v}_j|,
\end{eqnarray}
where $\vec{r}_i$ and $\vec{v}_i$ is the position and velocity of the $i$th test particle, respectively.
The dispersities of relative azimuthal angle of position and velocity of $N$ test particles, are defined by,
\begin{eqnarray}\label{azimuthal angle}
	\left \langle D\cos{\Delta\phi}\right \rangle  &=&\frac{-1}{N(N-1)} \sum ^N _j \sum ^N _{i\neq j}\cos{(\phi_i-\phi_j)}, \notag \\
	\left \langle D\cos{\Delta\phi_v}\right \rangle  &=&\frac{-1}{ N(N-1) } \sum ^N _j \sum ^N _{i\neq j}\cos{(\phi_{v_i}-\phi_{v_j})},
\end{eqnarray}
where $\phi_i$ and $\phi_{v_i}$ are the azimuthal angles of position and velocity of the $i$-th test particle, respectively. These defined observables are sensitive to the dispersity status of test particles. 

On the other hand, the normalized position and velocity are defined as
\textit{}\begin{eqnarray}\label{Normalized}
\left \langle N_r \right \rangle  &=& \frac{\sum ^N _i \vec{r}_{i}}{\sum ^N _i |\vec{r}_{i}|}, \notag \\
\left \langle N_v \right \rangle  &=& \frac{\sum ^N _i \vec{v}_{i}}{\sum ^N _i |\vec{v}_{i}|}.
\end{eqnarray}
At the same time, the averaged cosine values of the azimuthal angle of position and velocity are defined as
\begin{eqnarray}\label{Cosine}
	\left \langle \cos{\phi}\right \rangle  &=& \frac{1}{N}\sum ^N _i \cos{(\phi_{i}-\phi_i^{init})}, \notag \\
	\left \langle \cos{\phi_v}\right \rangle  &=& \frac{1}{N}\sum ^N _i \cos{(\phi_{v_i}-\phi_i^{init})},
\end{eqnarray}
where $\phi_i^{init}$ is the azimuthal angle of position of the $i$th test particle at $t_0$. We will take the event average for the above-defined observables. These observables are designed to measure the collectivity for the different configurations of the position and velocity of test particles. Table~\ref{table:obsopp} shows the expectations (i.e., event average values) of the eight defined observables for the initial states of cases 1 and 2. The expectations actually provide us the reference values for the limits of zero and strongest collectivity, since the initial state of case 1 corresponds to the status of zero collective flow, while the initial state of case 2 corresponds to the status of the strongest collective flow. Our aim is that by calculating the time evolutions of all these observables and comparing them with the reference values, we want to understand how collectivity is established or developed for cases 1 and 2 in large and small colliding systems.

\section{Results and discussions}
 \label{results}

 \subsection{Large colliding systems}
 \label{Au$+$Au}

\begin{figure}[htbp]
	\centering
	\includegraphics[width=0.9\linewidth]{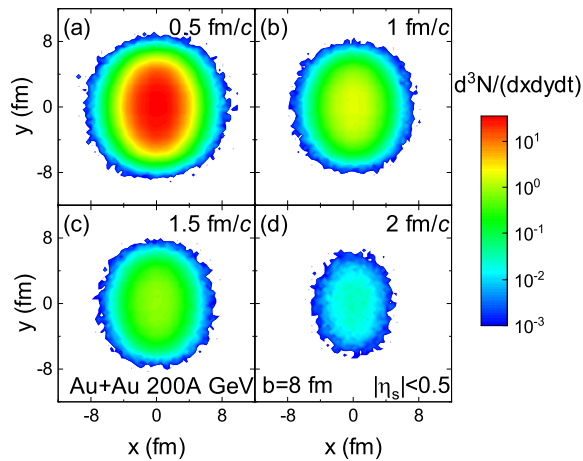}
	\caption{(Color	online) The distributions of initial partons ($|\eta_{s}|<$0.5) in the transverse plane at different times in Au $+$ Au midcentral collisions at 200 GeV from the AMPT model.}
	\label{Auparton}
\end{figure}

The two-dimensional (2D) distributions of initial partons (within a space-time rapidity window of $|\eta_{s}|<$0.5) in the transverse plane (the $x$-$y$ plane) at four selected times in Au$+$Au midcentral collisions at $\sqrt{s_{NN}}$ = 200 GeV ($b = 8$ fm) are shown in Fig.~\ref{Auparton}. The short axis of the fireball is along $x$ axis, and the long axis of the fireball is along $y$ axis. Most of the partons are generated near the center of the overlap region at the early time. Therefore, we add four test particles into the fireball which are placed at the position of $ \vec{r}$ = (3, 0, 0 fm) at the time of 1 fm/\textit{c} during the partonic stage, unless specified otherwise.  Then we will calculate the time evolutions of the above-defined observables to investigate how collective flow can be built or developed by parton collisions.

 \subsubsection{Case 1 for Au$+$Au midcentral collisions}
\begin{figure}[htbp]
	\centering
	\includegraphics[width=0.9\linewidth]{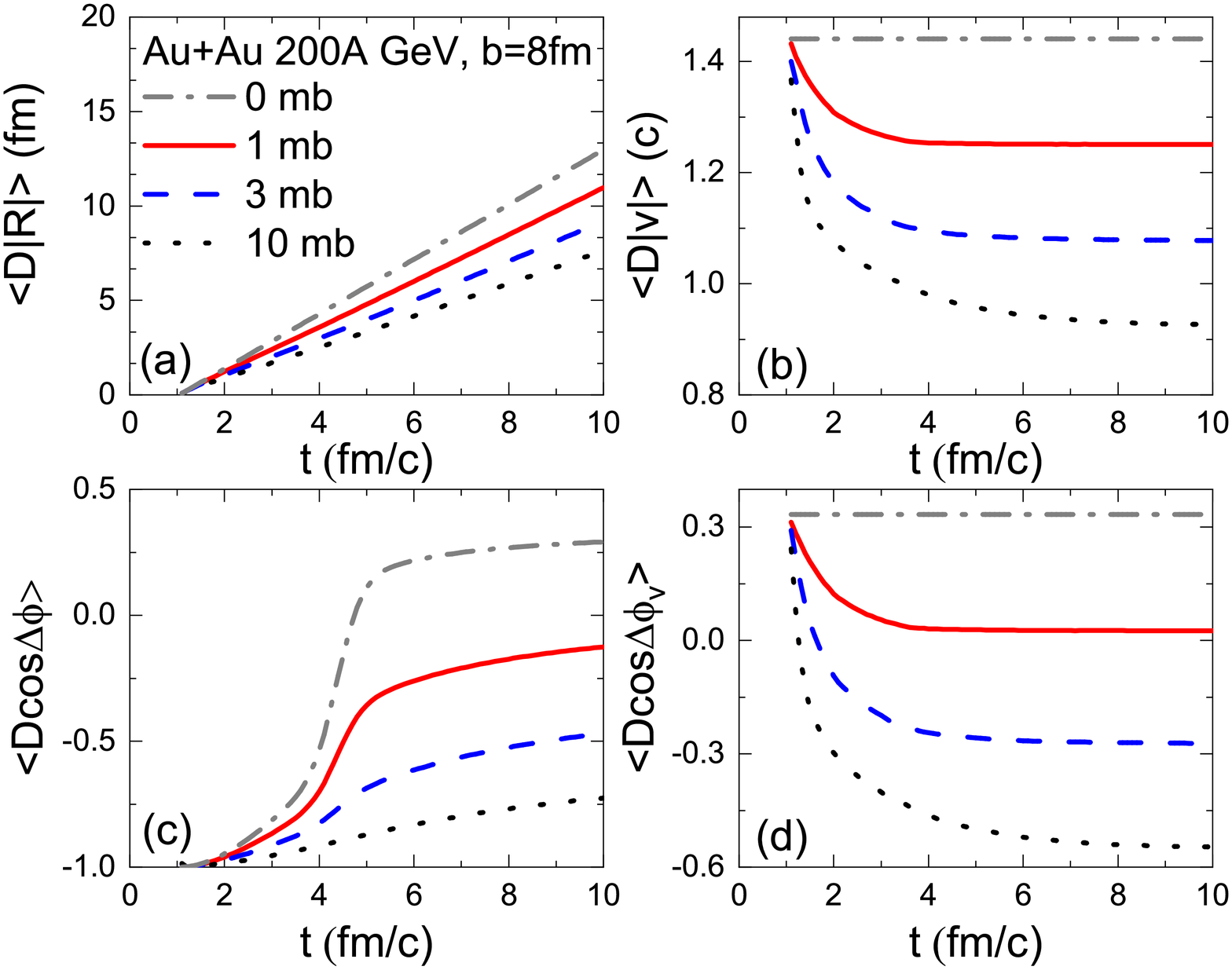}
	\caption{(Color	online) The time evolutions of the dispersities of four test particles for case 1 in Au$+$Au midcentral collisions at 200 GeV from the AMPT model with different parton cross sections.}
	\label{Auvxy0.9cdv}
\end{figure}

Figure~\ref{Auvxy0.9cdv} shows the time evolutions of the dispersities of the test particles with isotropic velocities for case 1 in Au$+$Au midcentral collisions at 200 GeV from the AMPT model with different parton interaction cross sections. The time evolutions of the dispersities in coordinate space are presented in Fig.~\ref{Auvxy0.9cdv}(a) and (c). The dispersities of position (and angle) increase with time but decrease with parton interaction cross section. It can be understood because if there is no parton collision, the test particles will keep their four different initial velocities and separate further and further over time, as illustrated by the results of 0 mb. However, our results show that if the test particles suffer more collisions with the medium, they are more difficult to separate in space. Considering Table~\ref{table:obsopp}, it is difficult to judge the strength of collectivity only by these two observables, since they give the same expectations for zero collectivity and the strongest collectivity. Figure~\ref{Auvxy0.9cdv}(b) and (d) show that dispersities of velocity and angle decrease and saturate with time, and they also decrease with parton interaction cross section. Considering Table~\ref{table:obsopp}, it indicates that the test particles tend to change their momentum towards a common direction through more parton collisions, thus acquiring collectivity. This supports the scenario that the test particles are affected by the stronger collective flow of the surrounding medium generated by the larger parton interaction cross section. In short, these behaviors of the dispersities of position and velocity suggest that although the test particles spread in space with time, they eventually tend to move in a common direction due to the collectivity of the surrounding medium built by parton collisions.

\begin{figure}[htbp]
	\centering
	\includegraphics[width=0.9\linewidth]{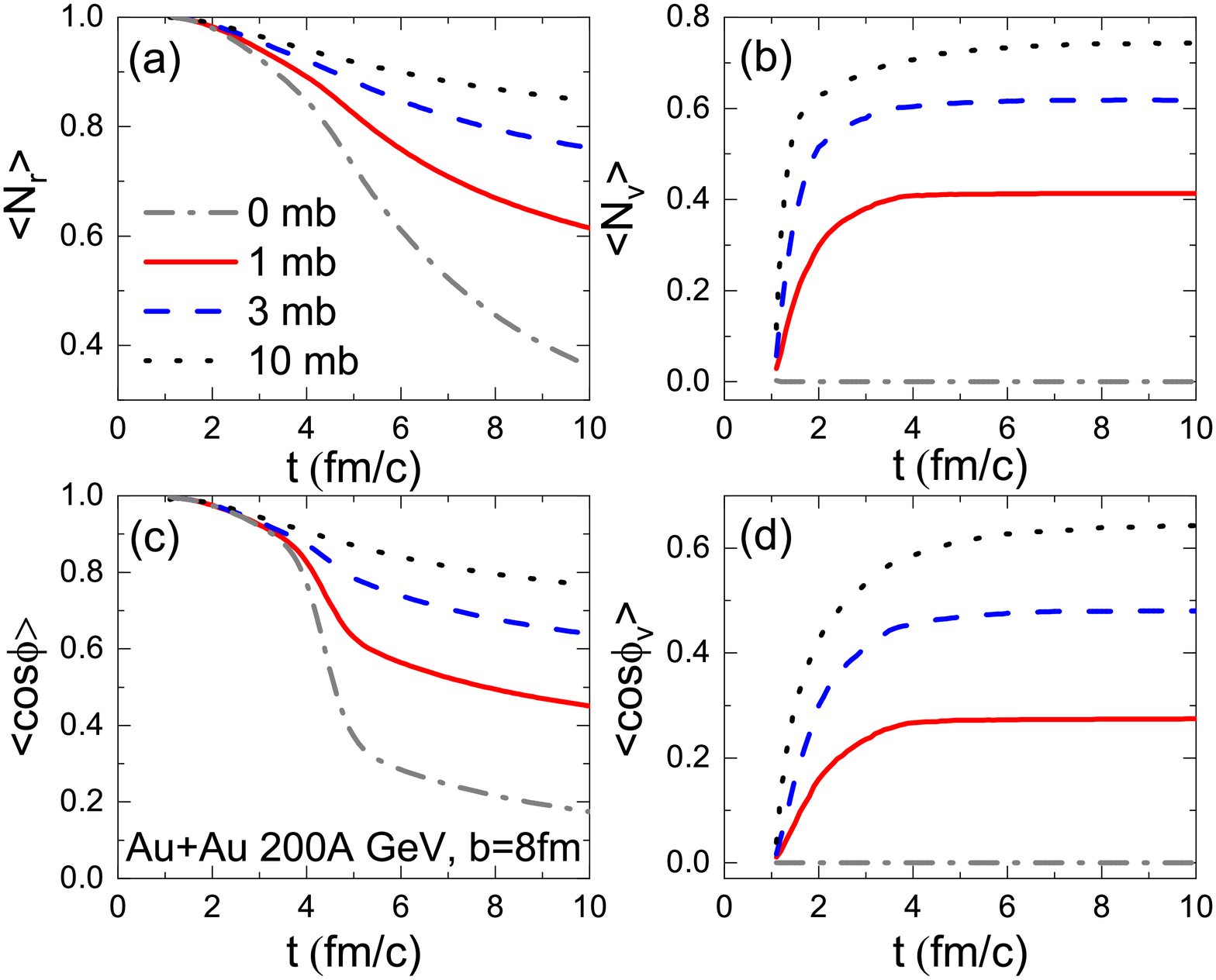}
	\caption{(Color	online) The time evolutions of (a) normalized position, (b) normalized velocity, (c) averaged cosine value of azimuthal angle of position, and (d) averaged cosine value of azimuthal angle of velocity of four test particles for case 1 in Au$+$Au midcentral collisions at 200 GeV from the AMPT model with different parton cross sections.}
	\label{Auvxy0.9cv}
\end{figure}

Figure~\ref{Auvxy0.9cv} shows the time evolutions of normalized positions and velocities, averaged cosine values of azimuthal angle of position and velocities of the test particles for case 1 in Au$+$Au midcentral collisions at 200 GeV from the AMPT model with different parton interaction cross sections. Figures~\ref{Auvxy0.9cv}(a) and (c) show that the normalized position and its mean cosine values of azimuthal angle decrease with time, but increase with parton interaction cross section. Considering Table~\ref{table:obsopp}, this also indicates that the test particles spread out with time, but more parton collisions prevent them from separating from each other further. On the other hand, Figs.~\ref{Auvxy0.9cv}(b) and (d) show that both the normalized velocities and averaged cosine values of azimuthal angle of velocities increase with time and parton interaction cross section. Considering Table~\ref{table:obsopp}, it indicates that the movement of the test particles changes from isotropic to collective, i.e., the test particles tend to move along the positive direction of the $x$ axis gradually. The trend becomes more significant with a larger parton interaction cross section, which reflects the formation of stronger collective flow due to more parton collisions.

\begin{figure}[htbp]
	\centering
	\includegraphics[width=0.9\linewidth]{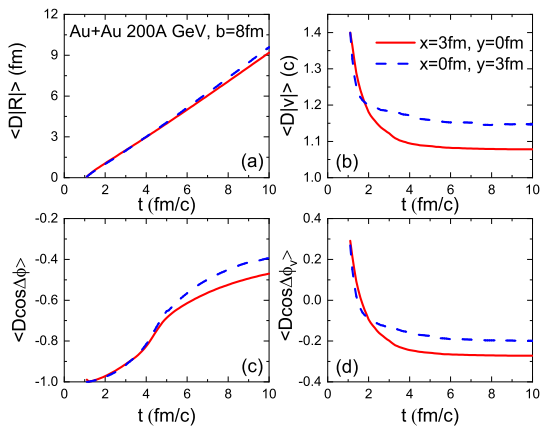}
	\caption{(Color	online) The time evolutions of the dispersities of four test particles for case 1 in Au$+$Au midcentral collisions at 200 GeV from the AMPT model with the parton cross section of 3 mb, where the test particles are placed at $\vec{r}_{0}$ = (3, 0, 0 fm) or (0, 3, 0 fm) at $t_0$ = 1 fm/\textit{c}.}
	\label{Auprosidv}
\end{figure}

If the collective flow is formed, is it anisotropic in Au$+$Au midcentral collisions? We can test the anisotropy of collectivity by placing the four test particles at $\vec{r}_{0}$ = (0,3,0 fm), instead of $\vec{r}_{0}$ = (3, 0, 0 fm) and make a comparison. Figure~\ref{Auprosidv} shows the comparisons of the time evolutions of the dispersities of four test particles for case 1 in Au$+$Au midcentral collisions at 200 GeV, between if the four test particles are placed at $\vec{r}_{0}$ = (3, 0, 0 fm) and if at $\vec{r}_{0}$ = (0, 3, 0 fm) at $t_0$ = 1 fm/\textit{c}. Considering Table~\ref{table:obsopp}, the four kinds of dispersities for $\vec{r}_{0}$ = (3, 0, 0 fm) are getting closer to the expectations for the strongest collectivity than those for $\vec{r}_{0}$ = (0, 3, 0 fm) at the end. This indicates that the collectivity formed along the $x$ axis is stronger than that along the $y$ axis.

\begin{figure}[htbp]
	\centering
	\includegraphics[width=0.9\linewidth]{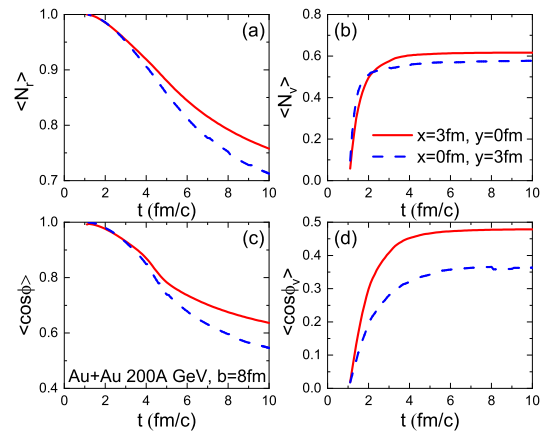}
	\caption{(Color	online) The time evolutions of (a) normalized position, (b) normalized velocity, (c) averaged cosine value of azimuthal angle of position, and (d) averaged cosine value of azimuthal angle of the velocity of four test particles for case 1 in Au$+$Au midcentral collisions at 200 GeV from the AMPT model with the parton cross section of 3 mb, where the test particles are placed at $\vec{r}_{0}$ = (3, 0, 0 fm) or (0, 3, 0 fm) at $t_0$ = 1 fm/\textit{c}. }
	\label{Auprosiv}
\end{figure}

Figure~\ref{Auprosiv} shows the comparisons of the time evolutions of normalized position, normalized velocity,  averaged cosine value of azimuthal angle of position, and averaged cosine value of azimuthal angle of the velocity of four test particles for case 1 in Au $+$ Au midcentral collisions at 200 GeV from the AMPT model, if the four test particles are placed at $\vec{r}_{0}$ = (3, 0, 0 fm) and if at $\vec{r}_{0}$ = (0, 3, 0 fm) at $t_0$ = 1 fm/\textit{c}. Similarly to Fig.~\ref{Auprosidv}, the four kinds of observables for $\vec{r}_{0}$ = (3, 0, 0 fm) are getting closer to the expectations for the strongest collectivity than those for $\vec{r}_{0}$ = (0, 3, 0 fm) at the end. In short, our results in Figs.~\ref{Auprosidv} and \ref{Auprosiv} indicate that the behaviors of test particles do reflect the elliptic anisotropy of collectivity in Au $+$ Au midcentral collisions at 200 GeV.

\subsubsection{Case 2 for Au$+$Au midcentral collisions}

\begin{figure}[htbp]
	\centering
	\includegraphics[width=0.9\linewidth]{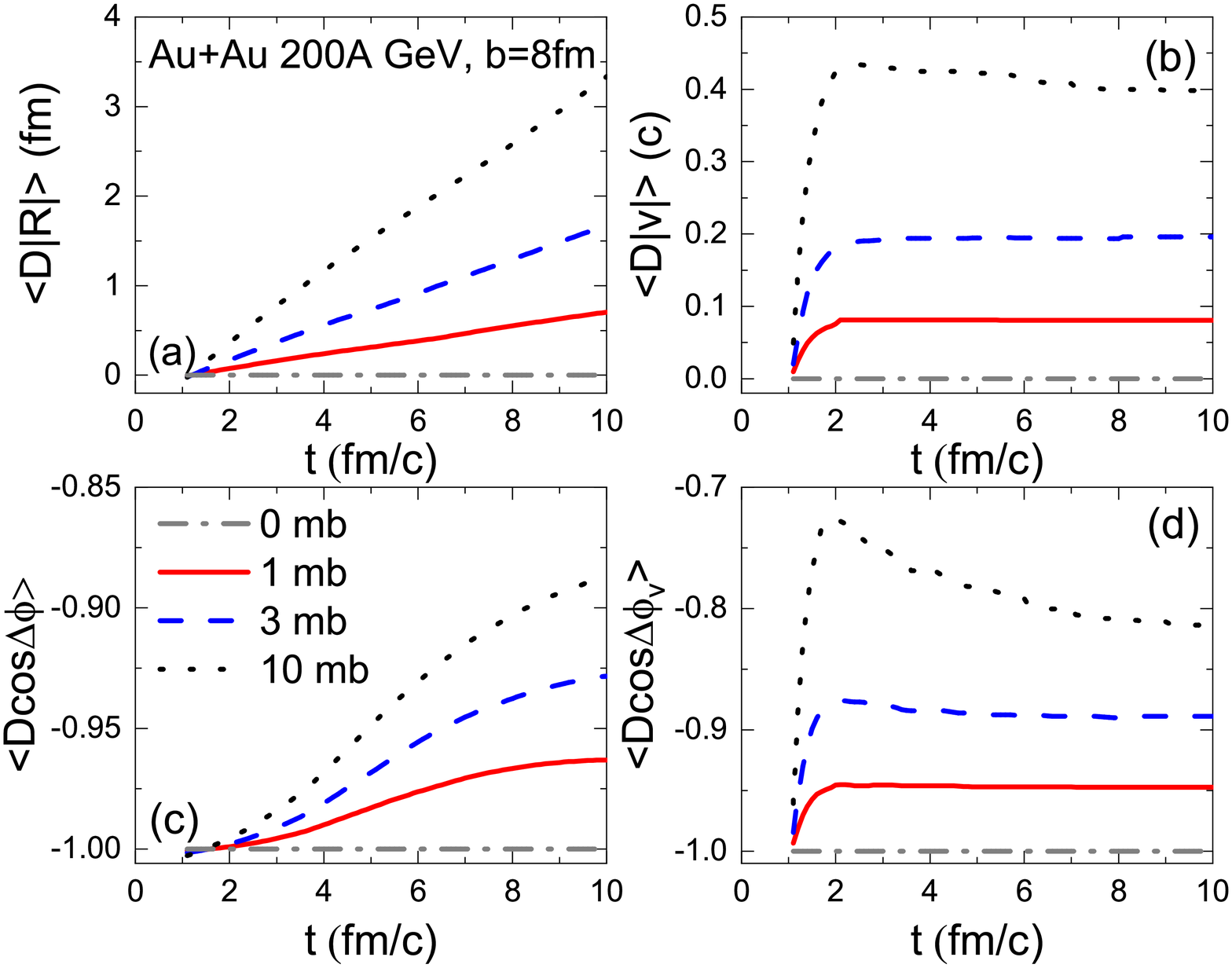}
	\caption{(Color	online) Same as Fig.~\ref{Auvxy0.9cdv}, but for case 2.}
	\label{Auvx0.9cdv}
\end{figure}

Figure~\ref{Auvx0.9cdv} shows the time evolutions of the dispersities of the test particles with a common initial velocity for case 2 in Au$+$Au midcentral collisions at 200 GeV from the AMPT model with different parton interaction cross sections. In Fig.~\ref{Auvx0.9cdv}(a) and (c), the dispersities of position (and angle) increase with time, which indicates that the test particles separate further and further over time. Opposite to case 1, the dispersities of position (and angle) increase with parton interaction cross section. It is because there is the strongest collectivity for the test particles in the initial state of case 2, however, the collectivity of test particles is more significantly damaged by more parton collisions. On the other hand, in Fig.~\ref{Auvx0.9cdv}(b) and (d), dispersities of velocity (and angle) increase with time, and they saturate for the small parton cross section but decrease for the large parton cross section.

\begin{figure}[htbp]
	\centering
	\includegraphics[width=0.9\linewidth]{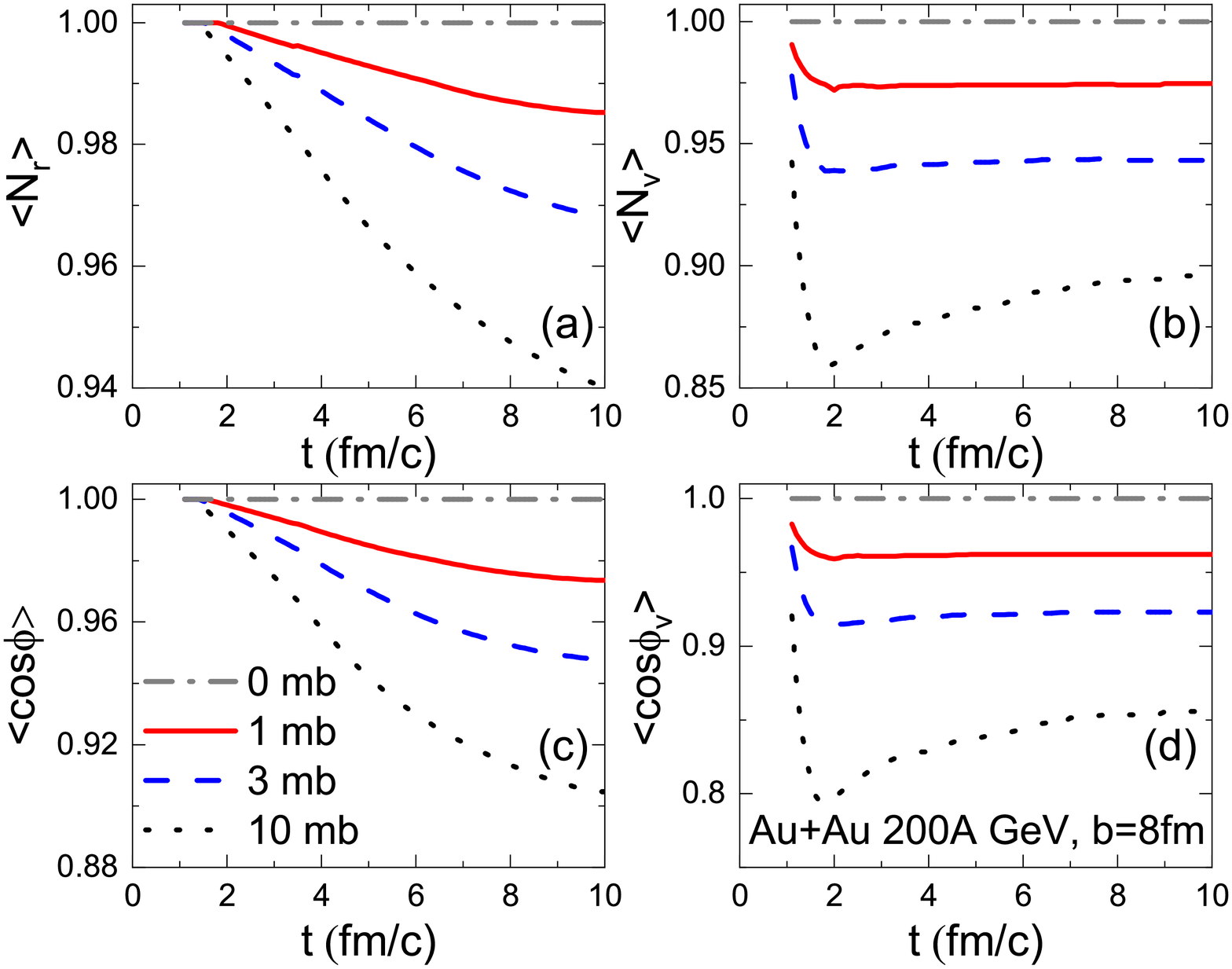}
	\caption{(Color	online) Same as Fig.~\ref{Auvxy0.9cv}, but for case 2.}
	\label{Auvx0.9cv}
\end{figure}

Figure.~\ref{Auvx0.9cv} shows the time evolutions of normalized positions and velocities, averaged cosine values of azimuthal angle of position, and velocities of the test particles for case 2 in Au$+$Au midcentral collisions at 200 GeV from the AMPT model with different parton interaction cross sections.  Consistent with Fig.~\ref{Auvx0.9cdv}(a) and (c), Fig.~\ref{Auvx0.9cv}(a) and (c) supports that the test particles separate further and further over time. On the other hand,  Fig.~\ref{Auvx0.9cv}(b) and (d) shows the normalized velocity and the averaged cosine value of azimuthal angle of velocity decrease at the beginning, and then rebound. The kink shape becomes more obvious as the parton interaction cross section increases.

\begin{figure}[htbp]
	\centering
	\includegraphics[width=0.9\linewidth]{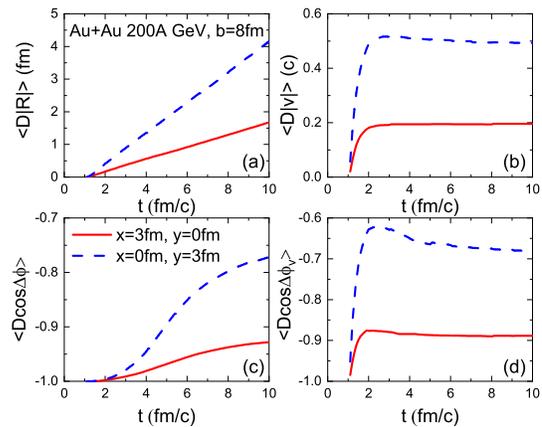}
	\caption{(Color	online) Same as Fig.~\ref{Auprosidv}, but for case 2.}
	\label{Auprosidvcase2}
\end{figure}

Figure.~\ref{Auprosidvcase2} shows the comparisons of the time evolutions of the dispersities of four test particles between if the test particles are placed at $\vec{r}_{0}$ = (3, 0, 0 fm) with a same initial velocity $\vec{v}=(0.9 c, 0, 0)$ and if at $\vec{r}_{0}$ = (0, 3, 0 fm) with a same initial velocity $\vec{v}=(0, 0.9 c, 0)$ at $t_0$ = 1 fm/\textit{c} for case 2 in Au$+$Au midcentral collisions at 200 GeV. Compared to the results for $\vec{r}_{0}$ = (3, 0, 0 fm) (solid curve), the four kinds of dispersities for $\vec{r}_{0}$ = (0, 3, 0 fm) are larger at the end, and show more obvious kink shapes. 

\begin{figure}[htbp]
	\centering
	\includegraphics[width=0.9\linewidth]{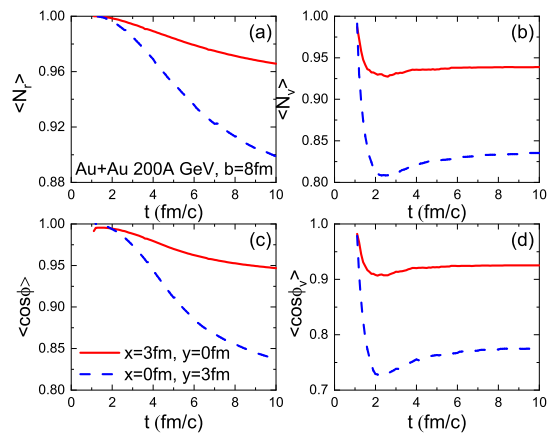}
	\caption{(Color	online) Same as Fig.~\ref{Auprosiv}, but for case 2.}
	\label{Auprosivcase2}
\end{figure}

Figure.~\ref{Auprosivcase2} shows the comparisons of the time evolutions of normalized position, normalized velocity,  averaged cosine value of azimuthal angle of position, and averaged cosine value of azimuthal angle of velocity of four test particles between the two same setting for Fig.~\ref{Auprosidvcase2} for case 2 in Au$+$Au midcentral collisions at 200 GeV. We also observe that the four kinds of observables from $\vec{r}_{0}$ = (0, 3, 0 fm) are lower than those from $\vec{r}_{0}$ = (3, 0, 0 fm), and show more obvious kink shapes, which is consistent with Fig.~\ref{Auprosidvcase2}.

\begin{figure}[htbp]
	\centering
	\includegraphics[width=0.9\linewidth]{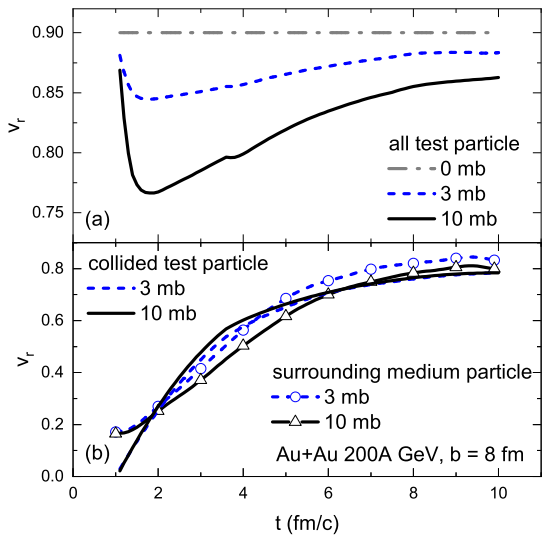}
	\caption{(Color	online) The time evolutions of the radial flow of (a) all test particles, (b) collided test particles, and surrounding medium particles for case 2 in Au$+$Au midcentral collisions at 200 GeV from the AMPT model with different parton cross sections, where the test particles are placed at $\vec{r}_{0}$ = (3, 0, 0 fm) at $t_0$ = 1 fm/\textit{c}.}
	\label{radialflow}
\end{figure}

To find out the source of the kink shapes for the above velocity-related observables, Fig.~\ref{radialflow} shows the time evolution of radial flow of all test particles, collided test particles, and surrounding medium particles for case 2 in Au$+$Au midcentral collisions at 200 GeV from the AMPT model with different parton cross sections. In our definition, the collided test particles are the test particles that have suffered collisions before the time $t$, while the surrounding medium particles are the medium particles that are near the test particles at the time $t$. In Fig.~\ref{radialflow}(a), we find that the shape of the time evolution of radial flow for all test particles is as same as the kink shape in Fig.~\ref{Auvx0.9cv}(d), which indicates that the radial flow should be responsible for the observed kink shape. To see the details, we compare the radial flow for collided test particles and surrounding medium particles in Fig.~\ref{radialflow}(b). In the beginning, the radial flow for collided test particles starts to increase, but is lower than that for surrounding medium particles.  At $t \sim 2$ fm/$c$ (the same time as the turning point in the above kink shapes), the radial flow for collided test particles catches up with that for surrounding medium particles. Then, the collided test particles show similar rising behavior to the surrounding medium particles, which indicates that they have become a part of medium particles and expanded with a common radial flow. Therefore, the kink structures in case 2 are actually the result of radial flow, and the test particles probe the effect of radial flow evolution in the partonic system.

\subsection{Small colliding systems}
\label{$p$ $+$ Pb}

\begin{figure}[htbp]
	\centering
	\includegraphics[width=0.9\linewidth]{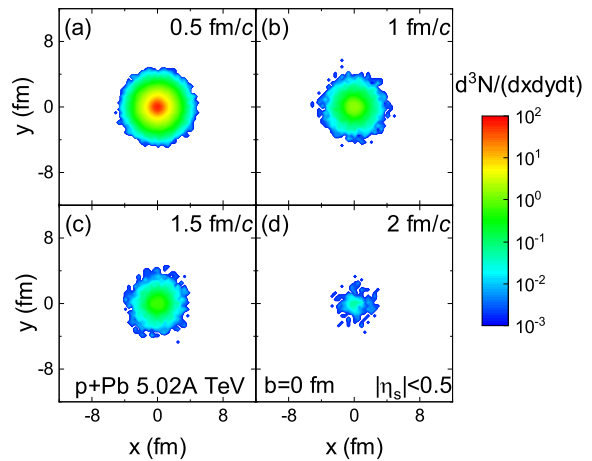}
	\caption{(Color	online) The distributions of initial partons ($|\eta_{s}|<$0.5) in the transverse plane at different times in $p$ $+$ Pb central collisions at 5.02 TeV from the AMPT model.
	}
	\label{Pbparton}
\end{figure}

The two-dimensional (2D) distributions of initial partons (within a space-time rapidity window of $|\eta_{s}|<$0.5) in the transverse plane at four selected times  in $p$ $+$ Pb central collisions at $\sqrt{s_{NN}}$ = 5.02 TeV ($b = 0$ fm) are shown in Fig.~\ref{Pbparton}. Similar to Fig.~\ref{Auparton} for midcentral Au$+$Au collisions, most of partons are generated near the center of the overlap region during early time for central $p$ $+$ Pb collisions. However, the volume for central $p$ $+$ Pb collisions is smaller than that for midcentral Au$+$Au collisions, and the gradient of parton spatial distribution looks more isotropic in $p$ $+$ Pb central collisions than in Au$+$Au collisions. We add four test particles into the small fireball which are placed at the position of $ \vec{r}=(2, 0, 0 fm)$ at the time of 1 fm/\textit{c} in the partonic stage of $p$ $+$ Pb central collisions. Then we will calculate and compare the time evolutions of the above defined observables between $p$ $+$ Pb central collisions at $\sqrt{s_{NN}}$ = 5.02 TeV and midcentral Au$+$Au collisions at $\sqrt{s_{NN}}$ = 200 GeV.

\subsubsection{Case 1 for $p$ $+$ Pb central collisions}

\begin{figure}[htbp]
	\centering
	\includegraphics[width=0.9\linewidth]{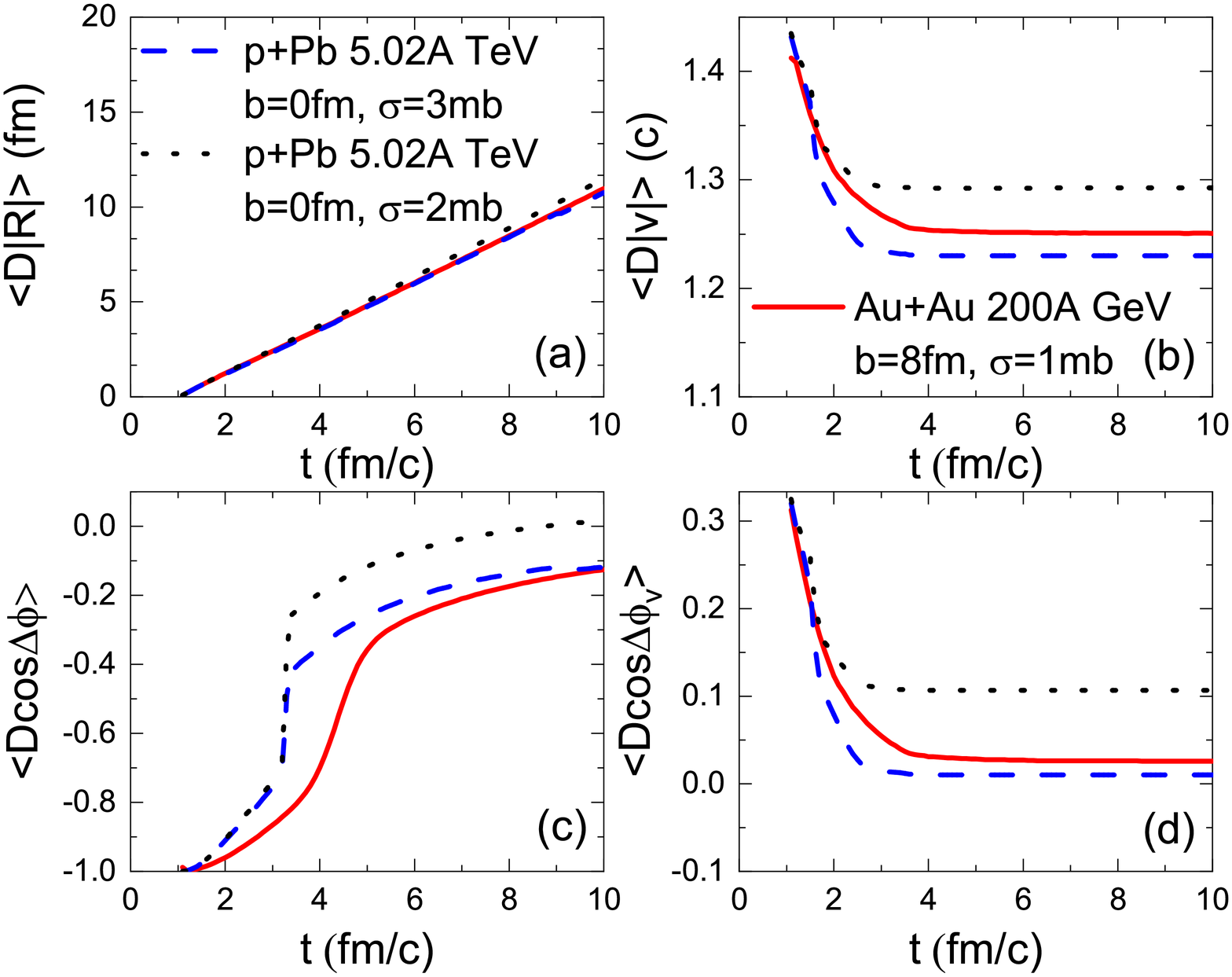}
	\caption{(Color	online) The time evolutions of the dispersities of four test particles for case 1 in $p$ $+$ Pb central collisions at 5.02 TeV from the AMPT model with a parton cross section $\sigma$ = 3 mb (dashed curve), and $\sigma$ = 2 mb (dotted curve), in comparisons with those for Au$+$Au midcentral collisions at 200 GeV with $\sigma$ = 1 mb (solid curve).}
	\label{Au_Pbpdvcase1}
\end{figure}

Figure.~\ref{Au_Pbpdvcase1} shows the time evolutions of the dispersities of the test particles with isotropic velocities for case 1 in $p$ $+$ Pb central collisions at 5.02 TeV from the AMPT model with two parton cross sections $\sigma$ = 3 mb and 2 mb. We find that $p$ $+$ Pb central collisions basically show similar trends to the results for Au$+$Au midcentral collisions at 200 GeV. However, the magnitudes for $p$ $+$ Pb results are very close to those for Au$+$Au midcentral collisions at 200 GeV with $\sigma$ = 1 mb.

\begin{figure}[htbp]
	\centering
	\includegraphics[width=0.9\linewidth]{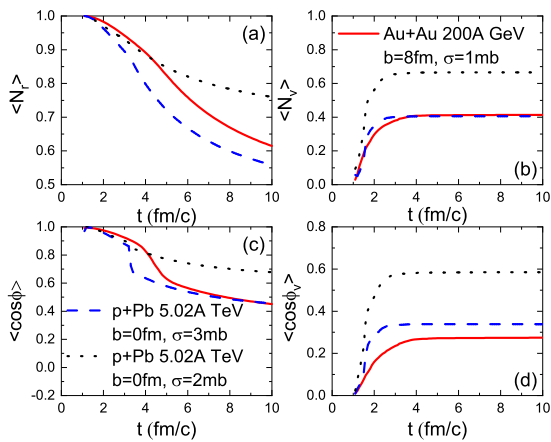}
	\caption{(Color	online) The time evolutions of (a) normalized position, (b) normalized velocity, (c) averaged cosine value of azimuthal angle of position, and (d) averaged cosine value of azimuthal angle of velocity of four test particles in $p$ $+$ Pb central collisions at 5.02 TeV from the AMPT model with a parton cross section $\sigma$ = 3 mb (dashed curve), and $\sigma$ = 2 mb (dotted curve), in comparisons with those for Au$+$Au midcentral collisions at 200 GeV with $\sigma$ = 1 mb (solid curve).}
	\label{Au_Pbpvcase1}
\end{figure}

Figure.~\ref{Au_Pbpvcase1} shows the time evolutions of normalized positions and velocities, averaged cosine values of azimuthal angle of position and velocities of the test particles for case 1 in $p$ $+$ Pb central collisions at 5.02 TeV from the AMPT model with two parton cross sections $\sigma =$ 3 mb and 2 mb. We find that $p$ $+$ Pb central collisions also show similar trends to the results for Au$+$Au midcentral collisions at 200 GeV.  We also observe that these observables for  $p$ $+$ Pb collisions are similar to Au$+$Au with $\sigma = 1$ mb. Since the parton interaction cross section of $\sim 3$ mb can well describe both large and small colliding systems at RHIC and the LHC, our results indicate that the collectivity in $p$ $+$ Pb central collisions at 5.02 TeV is much weaker than that in Au$+$Au midcentral collisions at 200 GeV. Notably, a system scan of small-size $A + A$ collisions should be very helpful in understanding the emergence of collectivity~\cite{Huang:2019tgz}(or nuclear structure~\cite{Ma:2022dbh,Li:2020vrg}) from small to large colliding systems.

\subsubsection{Case 2 for $p$ $+$ Pb central collisions}

\begin{figure}[htbp]
	\centering
	\includegraphics[width=0.9\linewidth]{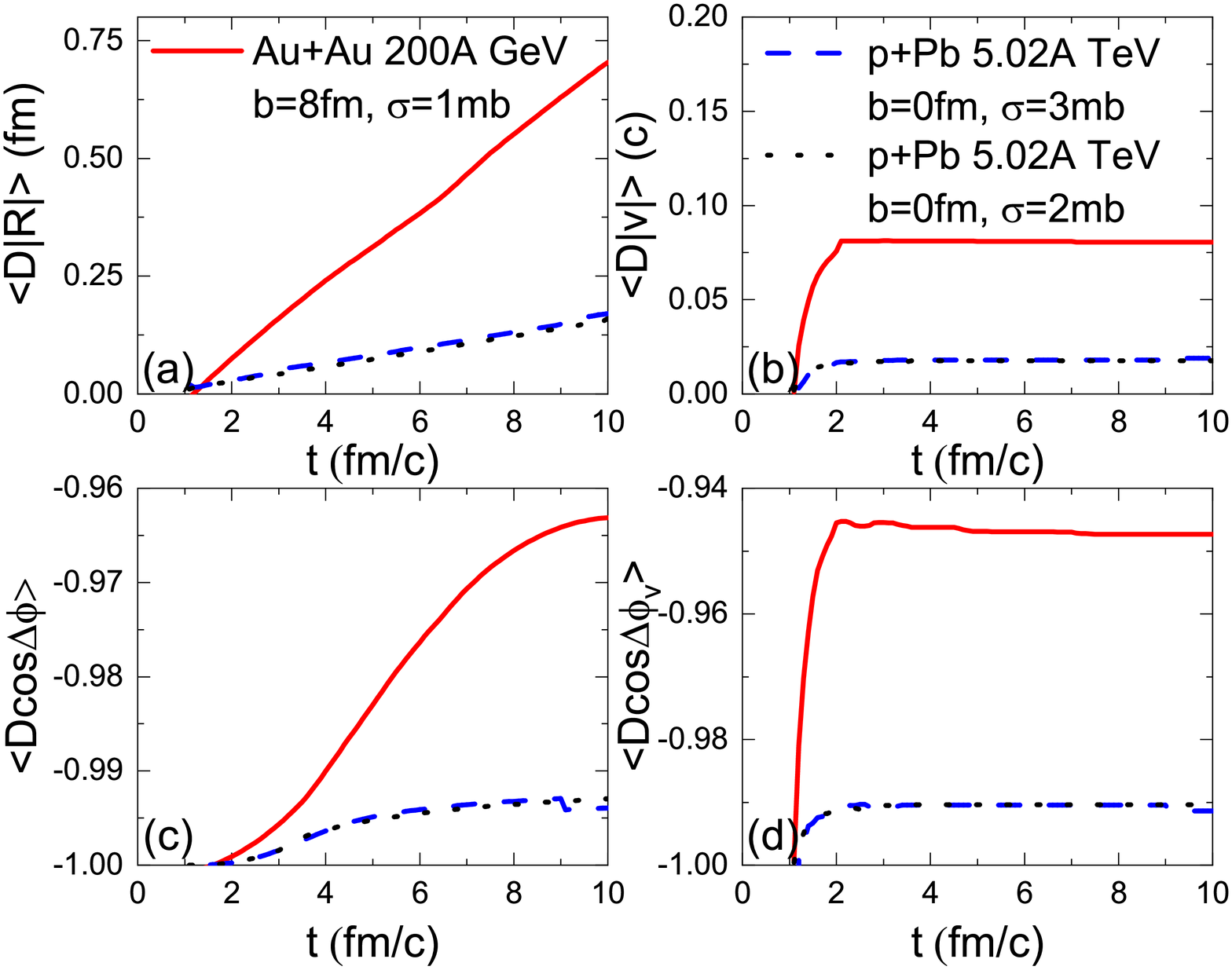}
	\caption{(Color	online)  Same as Fig.~\ref{Au_Pbpdvcase1}, but for case 2.}
	\label{Au_Pbpdvcase2}
\end{figure}

Figure.~\ref{Au_Pbpdvcase2} shows the time evolutions of the dispersities of the test particles with a common initial velocity for case 2 in $p$ $+$ Pb central collisions at 5.02 TeV from the AMPT model with two parton cross sections $\sigma$ = 3 mb and 2 mb. We observe that the dispersities of the test particles increase with time and are almost independent of parton interaction cross section. And the dispersities of the test particles in $p$ $+$ Pb central collisions are much less than those in Au $+$ Au midcentral collisions at 200 GeV with $\sigma$ = 1 mb.

\begin{figure}[htbp]
	\centering
	\includegraphics[width=0.9\linewidth]{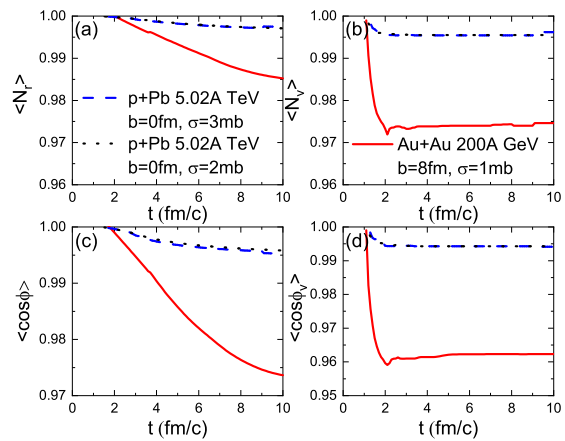}
	\caption{(Color	online) Same as Fig.~\ref{Au_Pbpvcase1}, but for case 2. }
	\label{Au_Pbpvcase2}
\end{figure}

Figure.~\ref{Au_Pbpvcase2} shows the time evolutions of normalized positions and velocities, averaged cosine values of azimuthal angle of position and velocities of the test particles for case 2 in $p$ $+$ Pb central collisions at 5.02 TeV from the AMPT model with two parton cross sections $\sigma$ = 3 mb and 2 mb. These observables decrease with time and are almost independent of parton interaction cross section for $p$ $+$ Pb central collisions at 5.02 TeV. The magnitudes of observables for $p$ $+$ Pb central collisions are much less than those for Au $+$ Au midcentral collisions. In contrast to Au$+$Au midcentral collisions, we observed only slight damage of the initial collective motion and insignificant kink structures in $p$ $+$ Pb central collisions, suggesting a lack of parton collisions in small colliding systems relative to large colliding systems. We observe more obvious differences in the fate of initial flow between small and large systems for case 2. This suggests that the ``flow'' from initial state correlation, e.g., CGC,  may be more likely to persist or survive in small colliding systems than in large colliding systems~\cite{Nie:2019swk, Greif:2017bnr, Schenke:2019pmk}.

\subsection{Four-test particle correlations in large and small colliding systems}
\label{cumulant}

To effectively reduce two-body nonflow contribution, multiparticle azimuthal cumulants have been proposed to
explore the collective flow of many-body systems \cite{Borghini:2000sa}. Similarly, we can define the first order of two-test particle and four-test particle azimuthal cumulants, as follows,
\begin{eqnarray}\label{cumulants}
c_{1}\left \{ 2 \right \} &=& \left \langle e^{i\left ( \phi_1-\phi_2 \right )} \right \rangle , \notag \\
c_{1}\left \{ 4 \right \} &=& \left \langle e^{i\left ( \phi_1+\phi_2-\phi_3-\phi_4 \right )} \right \rangle - 2 \left \langle e^{i\left ( \phi_1-\phi_2 \right )} \right \rangle ^2,
\end{eqnarray}
where $\phi_i$ is the azimuthal angle of the $i$th particle's transverse momentum. We expect that four-test particle cumulant can reflect the collectivity more effectively than two-test particle cumulant, since it can reduce few-body nonflow contribution.

\begin{figure}[htbp]
	\centering
	\includegraphics[width=0.9\linewidth]{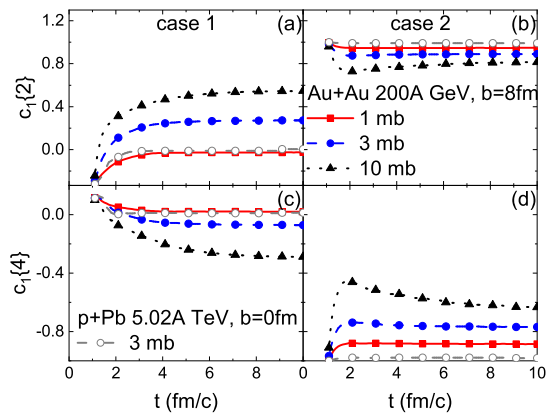}
	\caption{(Color	online) The time evolutions of the first order of two-test particle (top row) and four-test particle (bottom row) azimuthal cumulants in Au$+$Au midcentral collisions at 200 GeV and $p$ $+$ Pb central collisions at 5.02 TeV from the AMPT model with different parton interaction cross sections for cases 1 (left column) and 2 (right column).}
	\label{cumulants_f}
\end{figure}

Figure.~\ref{cumulants_f} shows that the time evolutions of first order of two-test particle $c_{1}\left\{ 2 \right \}$ and four-test particle $c_{1}\left\{ 4 \right \}$ azimuthal cumulants in Au$+$Au midcentral collisions at 200 GeV and $p$ $+$ Pb central collisions at 5.02 TeV from the AMPT model with different parton interaction cross sections for cases 1 and 2. Due to the definitions, we can easily see that the first order of two-particle azimuthal cumulant $c_{1}\left \{ 2 \right \}$ is actually the additive inverse of the above observable, i.e., the dispersity of relative azimuthal angle $\left \langle D\cos{\Delta\phi_v}\right \rangle$. Let us only focus on four-test particle azimuthal cumulant $c_{1}\left\{ 4 \right \}$ shown in the plot (c) and (d) for cases 1 and 2, respectively. In Fig.~\ref{cumulants_f}(c), for case 1, we can see that the four-test particle azimuthal cumulant $c_{1}\left\{ 4 \right \}$ becomes more negative with time and parton interaction cross section. It indicates that collective flow is built up by more and more parton collisions with time. It supports that more partonic collisions create stronger collective flow. On the other hand, Fig.~\ref{cumulants_f}(d) shows that case 2 has an opposite trend to case 1, where the four-test particle azimuthal cumulant becomes increasingly less negative with time and parton interaction cross section. It indicates that more parton collisions damage the initial collective flow more significantly. At the same time, we also observe that the four-test particle azimuthal cumulant first increases and then decreases with time for the large parton interaction cross section in Au$+$Au midcentral collisions. The behavior is consistent with that for two-test particle azimuthal cumulant $c_{1}\left\{ 2 \right \}$, which indicates that the collective motion is first damaged and then affected by radial flow due to a large number of parton collisions in Au$+$Au midcentral collisions. In terms of four-test particle azimuthal cumulants in small colliding systems, for case 1, the built collectivity in $p$ $+$ Pb central collisions at 5.02 TeV with $\sigma$ = 3 mb is just comparable to that with $\sigma$ = 1 mb in Au$+$Au midcentral collisions. For case 2, the four-test particle azimuthal cumulant in small colliding systems remains almost as negative as the initial value due to the small number of parton collisions in small colliding systems. Thus, most of the initial collectivity remains for case 2 in small colliding systems.

\section{Summary}
\label{summary}

We propose a test-particle method to investigate the parton collisional effects on the collective flow of partonic matter created in large and small colliding systems at RHIC and the LHC. We place test particles as passengers into the carrier medium created by Au$+$Au midcentral collisions at $\sqrt{s_{NN}}$ = 200 GeV and $p$ $+$ Pb central collisions at $\sqrt{s_{NN}}$ = 5.02 TeV. To study the transport dynamic of establishment and development of collective flow, we focus on two extreme cases corresponding to without any initial flow (case 1) and with an initial flow (case 2) of test particles. With case 1, we find that parton collisions play a significant role to build up the collectivity in both large and small colliding systems. Compared to small colliding systems, much stronger collectivity can be built up by parton collisions in large colliding systems, because more parton collisions build up stronger collectivity. With case 2, we find that the initial collectivity of test particles is more significantly damaged and affected by radial flow because of more parton collisions in large colliding systems, relative to small colliding systems. It suggests that the collectivity from the initial state can persist or survive more easily in small colliding systems than in large colliding systems. Our study provides a new method to understand the origin of collectivity in large and small colliding systems at RHIC and the LHC.

\begin{acknowledgments}

We thank Dr. Adam Bzdak for helpful discussions. This work is supported by the National Natural Science Foundation of China under Grants No. 11961131011, 12147101,  11890714,  11835002,  11421505, the Strategic Priority Research Program of Chinese Academy of Sciences under Grant No. XDB34030000, the Guangdong Major Project of Basic and Applied Basic Research under Grant No. 2020B0301030008, and the National Key Research and Development Program of China under Grant No. 2022YFA1604900.

\end{acknowledgments}

\end{CJK*}

\end{document}